\documentclass[twocolumn,superscriptaddress,floatfix,prb,amsmath,aps,showpacs]{revtex4}
\usepackage{graphicx}
\usepackage{bm}
\usepackage{amssymb}
\usepackage[dvips]{color}
\begin{document}
\bibliographystyle{apsrev}
\def\nn{\nonumber}
\def\dag{\dagger}
\def\u{\uparrow}
\def\d{\downarrow}
\def\j{\bm j}
\def\m{\bm m}
\def\l{\bm l}
\def\0{\bm 0}
\def\k{\bm k}

\title{\textbf{Nature of Possible Magnetic Phases 
in Frustrated Hyperkagome Iridate}}

\author{Ryuichi Shindou}
\email{rshindou@pku.edu.cn}
\affiliation{International Center for Quantum Materials, Peking University, Beijing, China}
\affiliation{Collaborative Innovation Center of Quantum Matter, Beijing, China}

\begin{abstract}
Based on Kitaev-Heisenberg model with 
Dzyaloshinskii-Moriya (DM) interactions, we studied nature of  
possible magnetic phases in frustrated hyperkagome iridate, 
Na$_4$Ir$_{3}$O$_8$ (Na-438).  Using Monte-Carlo simulation, we showed that 
the phase diagram is mostly covered by two competing 
magnetic ordered phases; Z$_2$ symmetry breaking (SB) phase and 
Z$_6$ SB phase, latter of which is stabilized by the classical order by disorder.  
These two phases are intervened by a first order phase transition
line with Z$_8$-like symmetry. 
The critical nature at the Z$_6$ SB ordering temperature 
is characterized by the 3D XY universality class, below which 
U(1) to Z$_6$ crossover phenomena appears; the Z$_6$ spin anisotropy 
becomes irrelevant in a length scale shorter than a crossover length 
$\Lambda_*$ while becomes relevant otherwise. 
A possible phenomenology of polycrystalline 
Na-438 is discussed based on this crossover phenomena.   
\end{abstract}
\maketitle

\section{Introduction}
An interplay between spin and orbital degree of freedom enriches 
the physics of Mott insulators~\cite{kk}. A strong relativistic spin-orbit 
interaction in cubic IrO$_6$ octahedra endows Ir electron with 
spin-orbit entangled Kramers doublet (pseudo-spin doublet),~\cite{bjkim} 
which opens a new root to ``$J=\frac{1}{2}$'' Mott insulators with 
coordinate-dependent pseudo-spin-anisotropic exchange 
interactions.~\cite{jk,cjk,cb} Mott insulating 
materials with bond-dependent spin-anisotropic interactions possibly 
stabilize gapless quantum spin liquid state with exotic excitations 
such as Majorana fermions~\cite{kitaev}.  
Honeycomb-lattice Na$_2$IrO$_3$,~\cite{na-213-a,na-213-rixs,na-213-neutron-1,
na-213-neutron-2} 
$\alpha$, $\beta$, $\gamma$-Li$_2$IrO$_3$,~\cite{a-li-213-a,b-li-213-a,g-li-213-a} 
and hyperkagome-lattice Na$_4$Ir$_3$O$_8$ (Na-438)~\cite{ot,sg,dally,shockley} are 
being intensively explored along this research interest. 
Throughout thermodynamic measurements,~\cite{na-213-a,a-li-213-a,b-li-213-a,g-li-213-a} 
resonant inelastic X-ray~\cite{na-213-rixs,b-li-213-rixs,g-li-213-rixs} and 
neutron scattering experiments~\cite{na-213-neutron-1,na-213-neutron-2}, 
the nature of low-temperature magnetic phases in all the 
honeycomb iridate compounds has been mostly clarified. 
Meanwhile nature of a low-$T$ phase of Na-438 is still veiled in mystery, 
although possible magnetic,~\cite{cb,kimchi} nematic~\cite{lawler-1,zhitomirsky}, 
valence bond solid,~\cite{bergholtz} and spin liquid phases~\cite{lawler-2,zhou} are 
being proposed theoretically. Early thermodynamic measurements such as 
magnetic specific heat, heat capacity, and magnetic susceptibility indicate 
spin liquid feature in Na-438~\cite{ot,sg}, while recent neutron scattering and 
muon spin relaxation experiments on powder samples suggest possibilities  
of  a short-range ordered quasistatic spin state (dubbed as ``configurationally 
degenerate phases with fluctuating order'')  or spin-freezing like phase in Na-438
~\cite{dally,shockley}. One of the major obstacles toward comprehensive understanding 
of low-$T$ magnetic properties of Na-438 stems from its 
low crystal symmetry, which results in complexity of electronic band structure 
and effective exchange model of localized spins.~\cite{cb,norman,micklitz}        
   
In this paper, we introduce an effective spin exchange model for 
the hyperkagome iridate to obtain a comprehensive 
understanding of classical magnetism possible in Na-438.  
Based on a lattice parameter of Na-438, we postulate additional 
lattice symmetries other than an exact crystal symmetry, to derive a 
relatively simpler but realistic effective spin model (Sec.~II). Using Monte 
Carlo (MC) simulation and Luttinger-Tisza (LT) analysis, 
we derive a classical magnetic phase diagram for Na-438 (Sec.~III). 
The phase diagram is mostly covered by two phases; one is 
Z$_2$ symmetry breaking (SB) magnetic phase and the other 
is Z$_6$ SB magnetic phase. The Z$_6$ anisotropy in the latter 
phase is attributed to the entropy effect (Sec.~IV). At finite temperature, 
these two phases are separated by a first order phase transition 
line with Z$_8$ symmetry. The finite size scaling (FNS) analysis concludes 
that criticality at the ordering temperatures 
of Z$_2$ and Z$_6$ phases are characterized by the 3D Ising, and 3D XY 
universality class respectively. For a finite-size system, an 
intermediate temperature regime appears below the ordering 
temperature of Z$_6$ phases, where the Z$_6$ spin  
anisotropy becomes effectively irrelevant and spin ordering 
develops in a U(1) symmetric way. This crossover 
temperature regime $\Delta T_{*} \equiv T_c-T_{*}$ is scaled with the linear 
dimension of the system size $\Lambda$ as 
$\Delta T_{*} \simeq \Lambda^{-\frac{1}{\nu_6}}$ with  
$\nu_6 = 1.45 \sim 1.85$ (Sec. V). Effects of the quantum fluctuation 
are also mentioned in Sec.~VI. A possible phenomenology of 
low-$T$ magnetic behaviors of the powder samples as well as 
a brief summary are given in Sec. VII.

\section{an exchange spin model for Hyperkagome iridate}
\begin{figure}[t]
\begin{center}
\includegraphics[width=80mm]{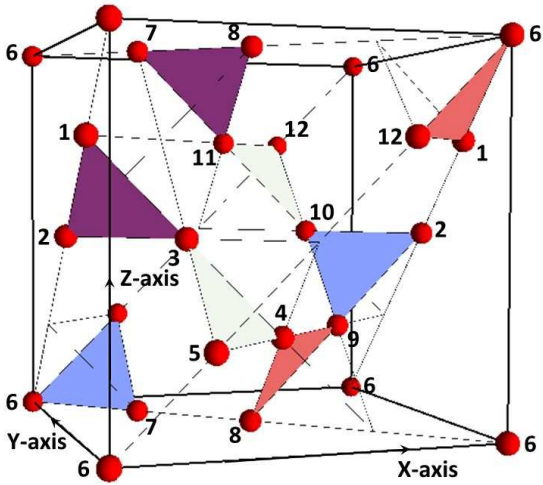}
 \caption{Cubic unit cell of the hyperkagome lattice, where red sphere 
 stands for iridium atom. The cubic unit cell contains 12 inequivalent iridium 
 sites.}
 \label{fig1}
 \end{center}
\end{figure}

The hyperkagome lattice is a three-dimensional lattice which comprises 
of corner-sharing triangles (Fig.~\ref{fig1}). 
The cubic unit cell contains 12 crystallographically distinct lattice points. Each sublattice 
point has a two-fold rotational axis, around which the lattice is symmetric under the $C_2$ rotation; 
$C_{2@j}$ ($j=1,2\cdots,12$) and $j$ sublattice index. We can choose 
$C_{2@j}$ ($j=1,\cdots,6$) as generators of the exact crystal symmetry group; the others six   
are identical to one of the generators  ($C_{2@7}=C_{2@2}$, $C_{2@8} = C_{2@1}$, $C_{2@9}=C_{2@6}$, 
$C_{2@10}=C_{2@5}$, $C_{2@11}=C_{2@3}$, $C_{2@12}=C_{2@4}$ in Fig.~\ref{fig1}). The two-fold rotational 
axes of $C_{2@1}$, $C_{2@2}$,$C_{2@3}$, $C_{2@4}$, $C_{2@5}$ and $C_{2@6}$ are 
along (0,1,1), (1,1,0), (1,0,1),  
(1,-1,0), (0,1,-1), and (1,0,-1) directions in the cubic unit cell.~\cite{ot,cb}.  
By these $C_2$ rotations, a form of the exchange interaction between a pair of 
nearest neighboring Ir pseudo-spin doublets determines all the others; 
\begin{widetext}
\begin{eqnarray}
H = \sum_{\langle i,j\rangle} \left(\begin{array}{ccc} 
S_{i,\mu} & S_{i,\nu} & S_{i,\rho} \\
\end{array}\right) \left(\begin{array}{ccc} 
J_1 & G_{12} + D_{12} & G_{31} + D_{31} \\ 
G_{12} - D_{12} & J_2 & G_{23} + D_{23} \\
G_{31} - D_{31} & G_{23} - D_{23} & J_3 \\
\end{array}\right) \left(\begin{array} {c}
S_{j,\mu} \\
S_{j,\nu} \\
S_{j,\rho} \\
\end{array}\right), \label{hami}
\end{eqnarray}
\end{widetext}
with  
\begin{align}
&\hspace{-0.5cm} 
\left(\begin{array}{ccc}
S_{i(j),\mu} & S_{i(j),\nu} & S_{i(j),\rho} \\
\end{array}\right) 
= \nn \\
& \left\{\begin{array}{cc}
\left(\begin{array}{ccc}
S_{i(j),x} & \pm S_{i(j),y} & S_{i(j),z} \\
\end{array}\right)  &  {\rm for} \!\  \!\ 
\langle i,j\rangle = \langle1,2\rangle, \langle 10, 12\rangle, \\
\left(\begin{array}{ccc}
S_{i(j),z} & \pm S_{i(j),x} & S_{i(j),y} \\
\end{array}\right)  & {\rm for} \!\ \!\ 
\langle i,j\rangle = \langle2,3\rangle, \langle 4, 9\rangle, \\ 
\left(\begin{array}{ccc}
S_{i(j),y} & \pm S_{i(j),z} & S_{i(j),x} \\
\end{array}\right)  & {\rm for} \!\ \!\  
\langle i,j\rangle = \langle3,1\rangle, \langle 6,5\rangle,  \\ 
\left(\begin{array}{ccc}
S_{i(j),z} & S_{i(j),y} & \pm S_{i(j),x} \\
\end{array}\right)  & {\rm for} \!\ \!\ 
\langle i,j\rangle = \langle7,8\rangle, \langle 12, 1\rangle, \\ 
\left(\begin{array}{ccc}
\pm S_{i(j),z} & \mp S_{i(j),y} & S_{i(j),x} \\
\end{array}\right)  & {\rm for} \!\  \!\ 
\langle i,j\rangle = \langle4,5\rangle, \langle 2, 10\rangle,  \\ 
\left(\begin{array}{ccc}
S_{i(j),y} & S_{i(j),x} & \mp S_{i(j),z} \\
\end{array}\right)  & {\rm for} \!\ \!\  
\langle i,j\rangle = \langle9,2\rangle, \langle 11, 7\rangle, \\ 
\left(\begin{array}{ccc}
S_{i(j),x} & \pm S_{i(j),z} & \mp S_{i(j),y} \\
\end{array}\right)  & {\rm for} \!\ \!\ 
\langle i,j\rangle = \langle5,3\rangle, \langle 10, 9\rangle,  \\ 
\left(\begin{array}{ccc}
\mp S_{i(j),y} & \pm S_{i(j),x} & S_{i(j),z} \\
\end{array}\right)  & {\rm for} \!\ \!\ 
\langle i,j\rangle = \langle3,4\rangle, \langle 6, 12\rangle, \\ 
\left(\begin{array}{ccc}
\mp S_{i(j),x} & S_{i(j),y} & \pm S_{i(j),z} \\
\end{array}\right)  & {\rm for} \!\ \!\  
\langle i,j\rangle = \langle8,4\rangle, \langle 5, 7\rangle,  \\ 
\left(\begin{array}{ccc}
\mp S_{i(j),z} & S_{i(j),x} & \pm S_{i(j),y} \\
\end{array}\right)  & {\rm for} \!\ \!\  
\langle i,j\rangle = \langle7,6\rangle, \langle 12, 11\rangle, \\ 
\left(\begin{array}{ccc}
\pm S_{i(j),y} & S_{i(j),z} & \mp S_{i(j),x} \\
\end{array}\right)  & {\rm for} \!\ \!\  
\langle i,j\rangle = \langle9,8\rangle, \langle 11, 10\rangle,  \\ 
\left(\begin{array}{ccc}
\mp S_{i(j),x} & S_{i(j),z} & S_{i(j),y} \\
\end{array}\right)  & {\rm for} \!\ \!\ 
\langle i,j\rangle = \langle1,6\rangle, \langle 8, 11\rangle, \nn \\ 
\end{array} \right.   
\end{align}
$i,j=1,\cdots,12$ the sublattice index and 
the summation of $\langle i,j\rangle$ is taken over all the nearest neighbor 
Ir sites. The exchange spin interaction takes the $3$ by $3$ real-valued 
matrix form, containing both symmetric ($J_i$ and $G_{ij}$) and 
antisymmetric component ($D_{ij}$). 

Quantitatively, spin anisotropy in hyperkagome 
iridate is determined by relative strength  
between the atomic spin-orbit interaction and non-cubic crystal field splitting 
among $t_{2g}$ orbitals.~\cite{cb} In the larger spin-orbit coupling case, 
$J_{\rm eff}=\frac{1}{2}$ doublet respects the cubic symmetry of the 
crystal field.~\cite{cb,bjkim,jk,cjk} To reduce a number of the spin-model parameters 
in eq.~(\ref{hami}), 
we will further postulate the cubic symmetry of the crystal field as well as  
several additional symmetries which are nearly respected 
by the lattice parameters of Na-438. 

\begin{figure}
 \begin{center}
\includegraphics[width=80mm]{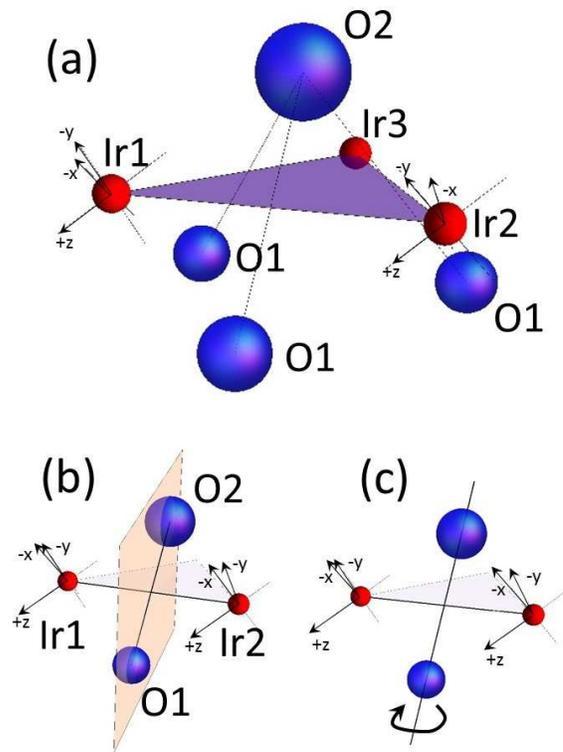}
 \caption{(a) A triangle formed by three neighboring iridium sites 
 (red spheres; Ir1, Ir2 and Ir3 in Fig.~\ref{fig1}) and their neighboring oxygens 
 (blue spheres; three O1 and one O2). O1 is shared by two of three oxygen   
 octahedra whose center accommodate Ir1, Ir2, and Ir3 respectively. 
 O2 is shared by the three oxygen octahedra. Cubic coordinates ($x$,$y$,$z$ coordinates) 
 with which an expression for the $J=\frac{1}{2}$ doublet, eq.~(\ref{J12}), is defined 
 are depicted by black solid lines with arrows at the iridium sites. 
 (b) Bond-centered mirror operation and (c) bond-centered $C_2$ rotation, 
 under which Ir1 and Ir2 are exchanged with each other, while O1 and O2 remain intact. 
 The lattice parameters of the hyperkagome iridate~\cite{ot} suggests that spatial 
 coordinates of O1, O2, Ir1, Ir2 and their cubic coordinates nearly 
 respect these two symmetries.}
 \label{fig2}
 \end{center}
 \end{figure}

To this end, note first that a form of the quadratic spin Hamiltonian between 
nearest neighboring Ir pseudo-spin doublets (for clarity 
of the explanation, choose Ir1 and Ir2 in Fig.~\ref{fig1} henceforth) is mostly 
determined by direct transfer integral between the 
two doublets and indirect transfer integrals  
mediated by oxygen $p$ orbitals neighboring to the two doublets 
(O1 and O2 in Fig.~\ref{fig2}). An electronic 
Hamiltonian for these four (Ir1, Ir2, O1 and O2 in Fig.~\ref{fig2}) 
is given by
\begin{widetext} 
\begin{align}
H^{\rm el} = & \epsilon_{{\rm Ir},d} 
\sum_{j={\rm Ir}1,{\rm Ir}2,\sigma} 
f^{\dagger}_{j,\sigma} f_{j,\sigma} +  
\sum_{m={\rm O}1,{\rm O}2,\sigma} \epsilon_{m,p} 
\sum_{a=x,y,z} p^{\dagger}_{m,a,\sigma} p_{m,a,\sigma} 
+ \sum_{i,j={\rm Ir}1,{\rm Ir}2,\sigma,\sigma'}
t_{ij,\sigma\sigma'} f^{\dagger}_{i,\sigma} f_{j,\sigma'} \nn \\
& + \sum_{i={\rm Ir}1,{\rm Ir}2,m={\rm O}1,{\rm O}2,\sigma,\sigma'} \Big[
s_{im,a,\sigma\sigma'} f^{\dagger}_{i,\sigma} p_{m,a,\sigma'} 
 + {\rm H.c.} \Big] + U_d \sum_{j={\rm Ir}1,{\rm Ir}2,\sigma,\sigma'} 
f^{\dagger}_{j,\sigma} f^{\dagger}_{j,\sigma'} f_{j,\sigma'} f_{j,\sigma}   \nn \\
& + \sum_{m={\rm O}1,{\rm O}2,a,b = x,y,z,\sigma,\sigma'}  U_{m,ab} \!\ \!\ \!\   
p^{\dagger}_{m,b,\sigma} p^{\dagger}_{m,a,\sigma'} p_{m,a,\sigma'} p_{m,b,\sigma} \nn \\
& +  \sum_{i={\rm Ir}1,{\rm Ir}2,m={\rm O}1,{\rm O}2,a = x,y,z,\sigma,\sigma'}  
U_{m,dp} \!\ \!\ \!\   
p^{\dagger}_{m,a,\sigma} f^{\dagger}_{i,\sigma'} f_{i,\sigma'} p_{m,a,\sigma}  \label{star}
\end{align} 
\end{widetext}
where $J_{\rm eff}=\frac12$ pseudo-spin doublet 
respects the cubic symmetry,~\cite{bjkim,jk,cjk,cb} 
\begin{eqnarray}
f_{j,\pm}= \frac{1}{\sqrt{3}} 
\Big\{\pm i d_{j,xz,s_{\frac{1}{2}}=\mp} + d_{j,yz,s_{\frac{1}{2}}=\mp} 
\pm  d_{j,xy,s_{\frac{1}{2}}=\pm}\Big\}.  \label{J12}
\end{eqnarray}
$\epsilon_{{\rm O}m,p}$ is an effective atomic energy for the oxygen O$m$ 
($m=1,2$)  and we assume that three $p$-orbitals at the respective  
oxygen ${\rm O}m$ take the same atomic energy, while 
$\epsilon_{{\rm O}1,p}\ne \epsilon_{{\rm O}2,p}$. 
$s_{im,a,\sigma\sigma'}$ stands for the transfer 
between the Ir doublet at $i=$ Ir1 or Ir2 and 
the neighboring oxygen $p$-orbital ($a=x,y,z$)  at $m=$ O1 or O2. 
$\sigma(\sigma')$ is the pseudo-spin index. 
$t_{ij,\sigma\sigma'}$ is the transfer between two nearest neighbor  
Ir doublets ($i,j=$Ir1 and Ir2). $U_d$, $U_{m,ab}$ and, $U_{m,dp}$ denote 
the on-site Coulomb interaction within Ir site, the on-site Coulomb interaction 
among three $p$ orbitals within O$m$ oxygen site and intersite Coulomb interaction 
between Ir site and O$m$ oxygen site respectively.  
Based on the strong coupling expansion, 
the exchange interaction between two Ir $J=\frac{1}{2}$ pseudospin 
doublets is derived from eq.~(\ref{star}). 

By way of $s_{im,a,\sigma\sigma'}$ and $t_{ij,\sigma\sigma'}$ in eq.~(\ref{star}), 
the form of the spin interaction between Ir1 and Ir2  
is constrained by those {\it additional} spatial symmetries {\it applied only for} Ir1, Ir2, O1 and O2. 
A lattice parameter of Na-438~\cite{ot} dictates that these four  
nearly respect two symmetries; one is a bond-centered mirror symmetry 
(Fig.~\ref{fig2}(b)) and the other is bond-centered $C_2$ rotational 
symmetry (Fig.~\ref{fig2}(c)). The mirror symmetry restricts the 
Dyzaloshinskii-Moriya (DM) vector to be in the mirror plane, while the $C_2$ 
rotation constrains the DM vector to be perpendicular to the $C_2$ rotational 
axis. When the two Ir doublets are chosen at Ir1 and Ir2 in Fig.~\ref{fig1}, the 
$C_2$ rotational axis is along $(0,1,1)$ and the mirror plane is perpendicular 
to the $(0,1,-1)$, so that the DM interaction vector between Ir1 and Ir2 
is along (1,0,0), i.e. $D_{12}=D_{31}=0$ in eq.~(\ref{hami}). 
Likewise, the symmetric part of the anisotropic exchange interaction 
between Ir1 and Ir2 respects $G_{12}=G_{31}=0$ and 
$J_2=J_3$ due to these two approximate symmetries. Without any justification, 
we further assume $G_{23}$ in eq.~(\ref{hami}) to be zero.  
This leads to the following reduced spin Hamiltonian for the hyperkagome iridate;     
\begin{eqnarray}
H = \sum_{\langle i,j\rangle} \left(\begin{array}{c} 
S_{i,\mu} \\ 
S_{i,\nu} \\ 
S_{i,\rho} \\ 
\end{array}\right)^T \left(\begin{array}{ccc} 
J+G & 0 & 0 \\ 
0 & J - G& D \\
0 & -D & J -G \\
\end{array}\right) \left(\begin{array} {c}
S_{j,\mu} \\
S_{j,\nu} \\
S_{j,\rho} \\
\end{array}\right),  \label{reduhami}
\end{eqnarray}
where $(S_{i(j),\mu} \!\ S_{i(j),\nu} \!\ S_{i(j),\rho})$ follows the same convention 
as above. This simplified spin model approximately 
includes all the effective spin exchange models previously 
derived.~\cite{norman,micklitz,cb} The actual values of $J$, $D$ and $G$ 
depend on detailed electronic band structure or assumptions.  
From the high-temperature expansion of eq.~(\ref{reduhami}), the 
Curie-Weiss temperature for polycrystalline samples is 
given by $3J-G$. From experimental Curie-Weiss fittings of the magnetic 
susceptibility of powder samples~\cite{ot,sg,dally}, we focus only on 
$3J>G$ region.  

\section{Monte Carlo Simulation}

\begin{figure}[htbp]
   \centering
   \includegraphics[width=80mm]{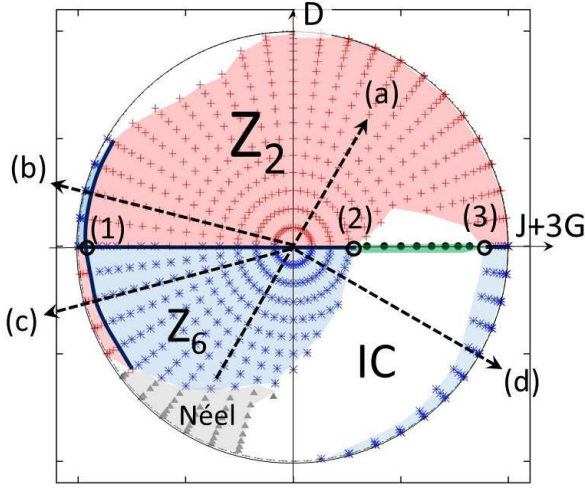} 
  \caption{Low-temperature magnetic phase diagram in $3J-G>0$ $(T=0.03)$. 
  Z$_2$ phase (red cross points) 
  Z$_6$ phases (blue double cross points), ferrimagnetic phase (green region with black 
  filled circle points), Neel phase (grey triangle points), and incommensurate (IC)
  magnetic phase (white region). Three open 
  circles at $D=0$ (labeled as (1,2,3)) denote high symmetric parameter points;  
  (1) a SU(2) point; $(J,D,G)=(0,0,-1)$, (2) AF Heisenberg point; 
  $(1,0,0)$, and (3) AF Kitaev point; $(\frac{1}{\sqrt{2}},0,\frac{1}{\sqrt{2}})$. 
  Four finite-$T$ phase diagrams in Fig.~\ref{fig4} (a,b,c,d) are along four arrows with black broken lines 
  labelled as (a,b,c,d) in Fig.~\ref{fig3} respectively.}
  \label{fig3} 
\end{figure}

Monte Carlo (MC) simulation was carried out for a $12\times L^3$ spin  
cluster of the hyperkagome lattice with periodic boundary condition ($L=6,\cdots,10$). 
We used the Metropolis method combined with an over-relaxation 
(microcanonical) update, where 1 MC step comprises of $12\times L^3$ 
number of single-spin flip trials followed by one or two (non-random) 
sequential applications of microcanonical update~\cite{creutz,zhitomirsky,kls}.  
To increase an acceptance rate of the single-spin flip trial, we impose maximum 
variation of spin to be smaller for lower temperature, $|\Delta {\bm S}| < T$ 
(the temperature unit is the square root of $J^2+D^2+G^2$). 
The microcanonical update comprises of a $\pi$-rotation of single spin around local 
exchange field created by its four neighboring spins. We apply the $\pi$-rotation 
on all the spins in sequence. After 30000  $\sim$ 40000 MC steps for 
the thermal equilibration, physical quantities are measured once per every 
5 MC steps. Physical quantities are typically averaged over 
20000 $\sim$ 40000 samples. 

\begin{figure}[t]
   \centering
   \includegraphics[width=80mm]{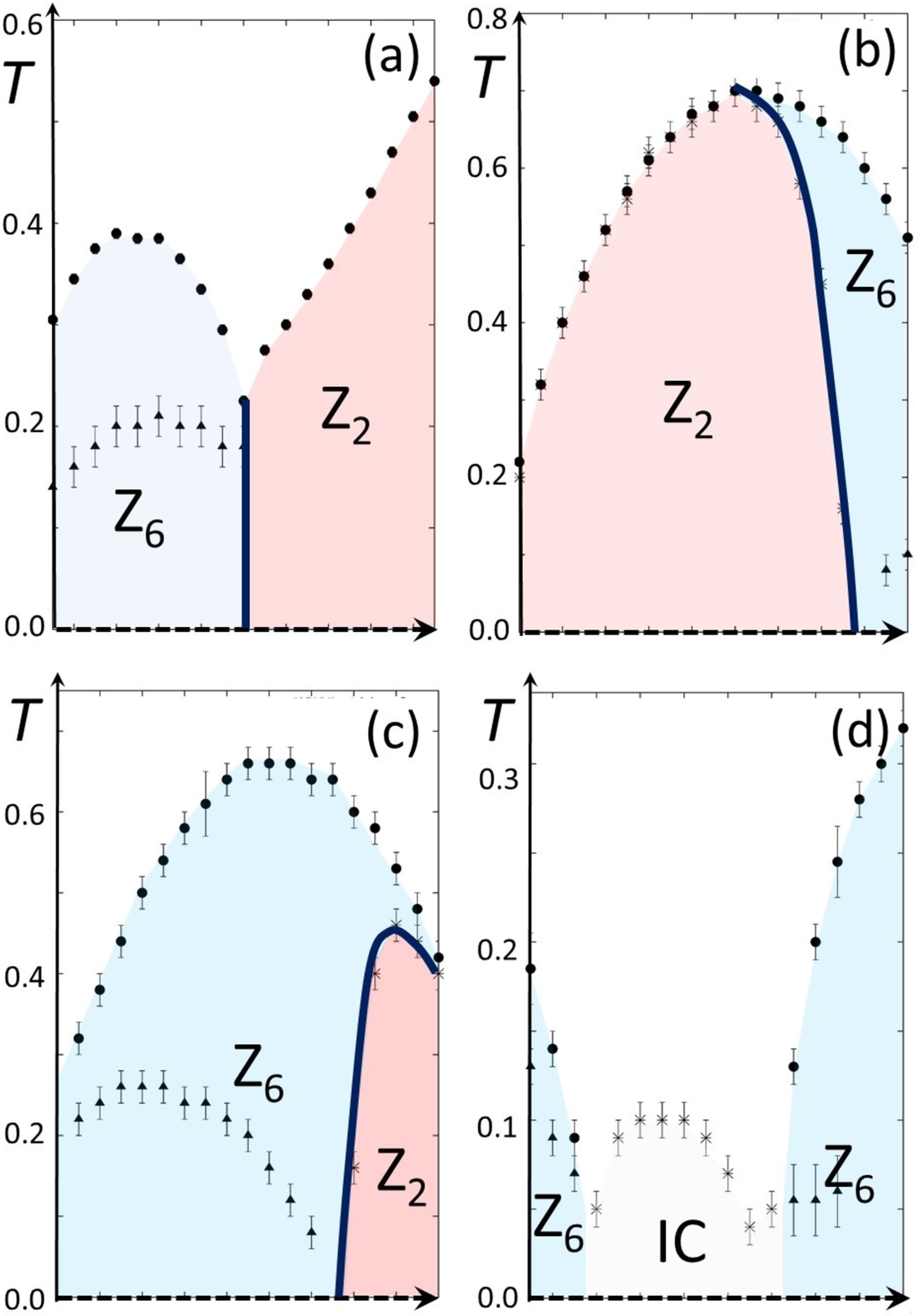}
  \caption{Finite-$T$ phase diagrams along the broken lines in 
  Fig.~\ref{fig3}. Critical temperatures ($T_c$) for Z$_2$, Z$_6$ SB phase and IC phase are all identified  
  with specific heat peak temperatures (black filled circle points). The crossover 
  temperature ($T_*$) within the Z$_6$ phase is depicted by black upper triangle points. 
  Phase boundary between Z$_6$ and Z$_2$ phases (depicted by bold purple lines) 
  is of the first order with Z$_8$-like symmetry (Appendix C). } 
  \label{fig4} 
\end{figure}

Fig.~\ref{fig3} and Fig. \ref{fig4} show low-$T$ and finite-$T$ magnetic 
phase diagrams obtained from the simulation. 
The low-$T$ diagram comprises of Z$_2$, Z$_6$, Z$_8$ non-coplanar antiferromagnetic phases 
and non-coplanar ferrimagnetic phase, all of which are associated with ${\bm k}=0$ 
ordering, Neel phase with ${\bm k}=(0,\pi,\pi)$ and incommensurate magnetic (IC) phase 
with ${\bm k}=(\alpha,\alpha,\alpha)$ with $0<\alpha<\pi$. The MC simulation does not  
see any magnetic orderings up to $T=0.03$ at the two high symmetry  
parameter points; antiferromagnetic (AF) Kitaev point ($J=G>0$, $D=0$) and 
isotropic AF Heisenberg point ($J>0$,$G=D=0$); see appendix 
A.~\cite{lawler-1,zhitomirsky} 

Z$_2$ and Z$_6$ SB magnetic phases and their stabilities can be captured  
by the Luttinger-Tisza (LT) analysis.~\cite{lt,luttinger} The analysis begins with 
the Fourier series of the quadratic spin Hamiltonian; 
\begin{eqnarray}
H = \sum [{\bm H}({\bm k})]_{(j,\alpha|m,\beta)} 
S_{j,\alpha}({\bm k}) S_{m,\beta}(-{\bm k}) \label{LT1}
\end{eqnarray} 
with $j,m$ sublattice index ($j,m=1,\cdots,12$) and $\alpha,\beta$ spin index.
Spin ordering in a magnetic ground state is specified by a lowest eigenmode of 
the $36$ by $36$ Hermitian matrix ${\bm H}({\bm k})$, provided that the eigenmode is 
real-valued and satisfies a fixed norm condition. The fixed norm condition requires 
that the norm over spin index is same for different sublattices. A magnetic 
structure of the Z$_2$ phase is given by  
the lowest eigenmode of ${\bm H}({\bm k})$ at ${\bm k}=0$, in which spin 
moment is ordered transverse to the $C_2$
rotational axis at respective site. The Z$_2$ magnetic phase is essentially 
same as the so-called `canted windmil phase' discussed in a previous work~\cite{cb}. 

\begin{figure}[t]
   \centering
   \includegraphics[width=86mm]{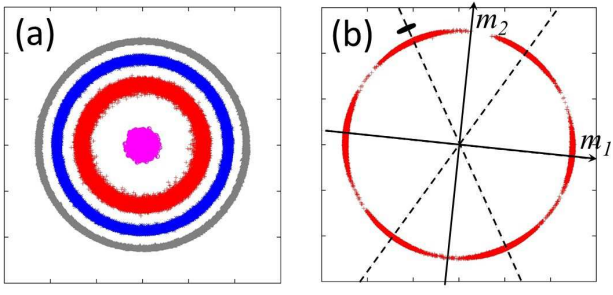} 
  \caption{Distribution of $(m_1,m_2)$ measured at different temperature and at 
  $(J,D,G)=(0.50,-0.22,-0.84)$ with $L=9$. (a) $T>T_{*}$ 
  ($T=0.8$ (pink), $0.64$ (red), $0.52$ (blue), $0.36$ (grey) from center to outside 
  while $T_c=0.691$). 
  (b) $T<T_{*}$ ($T=0.24$ (red), $0.04$ (black) from 
  center to outside). Six dotted lines with the Z$_6$ symmetry in (b) 
  specify the easy-axis directions within the 2D plane. We choose 
  $(J,D,G)=(0.50,-0.22,-0.84)$, because this point is proximate to the possible 
  candidate parameter point for Na-438 \cite{cb}.}
  \label{fig5} 
\end{figure}

The Z$_6$ SB magnetic phase is characterized by the doubly degenerate 
eigenmodes at ${\bm k}=0$ ($|\phi_1\rangle$ and $|\phi_2\rangle$), which 
form the 2D irreducible representation of ${\bm H}({\bm k}=0)$.  Any 
linear combination of these two cannot satisfy the fixed norm condition;  
moreover, they break the norm condition in the U(1) symmetric way.
Meanwhile the condition on averaged spin moment can be relaxed at finite 
temperature due to the thermal fluctuation. In fact, the doubly degenerate 
lowest eigenmodes of ${\bm H}({\bm k})$ appear at ${\bm k}=0$ in the 
Z$_6$ phase region, and spins become condensed into a plane 
subtended by these two below a critical temperature 
$T_c$ (determined by the specific heat peak). Fig.~\ref{fig5} 
shows a distribution of a projection of the 12 spins onto the 2D plane, 
i.e. $(m_1,m_2)$ defined by 
\begin{eqnarray}
m_{\mu} \equiv \frac{1}{N} \sum_{j,\alpha} S_{j,\alpha} \langle j,\alpha |\phi_{\mu} \rangle, 
\label{LT2}
\end{eqnarray} 
where $j$ is the site index and $\alpha$ is the spin index.  
For $T>T_c$, the projection is accumulated at origin 
($\langle \Psi \rangle=0$ with $\Psi \equiv m_1+i\!\ m_2$). For $T<T_c$, 
the amplitude develops continuously ($\langle |\Psi| \rangle \ne 0$). The 
simulation on a finite-size system suggests another characteristic temperature 
$T_* (<T_c)$, above which the distribution has the U(1) symmetry ($\langle \theta\rangle$ indefinite 
with $\Psi \equiv |\Psi|e^{i\theta}$) but below which it acquires an additional 
Z$_6$ structure ($ \langle \cos 6\theta \rangle \ne 0$ with $\Psi \equiv |\Psi|e^{i\theta}$).   

\section{classical order by disorder mechanism}
The Z$_6$ anisotropy at the lower temperature ($T<T_*$) 
can be understood from the classical order by disorder mechanism. 
To see this, let us expand ${\bm H}({\bm k})$ near  
${\bm k}=0$ with respect to small ${\bm k}$ and 
derive an effective Hamiltonian in the basis of $|\phi_1\rangle$ and 
$|\phi_2\rangle$ using the ${\bm k}\cdot {\bm p}$ perturbation. 
On the second order in ${\bm k}$, the effective Hamiltonian 
in the 2D plane subtended by $|\phi_1\rangle$ and $|\phi_2\rangle$ is ; 
\begin{eqnarray}
{\bm H}^{2\times 2}_{\rm eff} ({\bm k}) = \epsilon_0 + \hat{\bm V}^{2\times 2}_1({\bm k}) 
+ \sum_{n\ne 0} \frac{{\bm t}_{0n}({\bm k}) \cdot {\bm t}_{n0}({\bm k})}{\epsilon_n-\epsilon_0}. \nn 
\end{eqnarray}
$\epsilon_0$ is the lowest eigenenergy of ${\bm H}({\bm k}=0)$ to which $|\phi_1\rangle$ 
and $|\phi_2\rangle$ belong to. $\epsilon_n$ is a higher eigenenergy of ${\bm H}({\bm k}=0)$. 
${\bm V}^{2 \times 2}_1 ({\bm k})$ is a 2 by 2 block of $\Delta {\bm H}({\bm k}) \equiv 
{\bm H}({\bm k})-{\bm H}({\bm k}=0)$ 
in the basis of $|\phi_1\rangle$ and $|\phi_2\rangle$; 
$[{\bm V}^{2 \times 2}_1 ({\bm k})]_{ij} \equiv \langle \phi_i |\Delta {\bm H}({\bm k})|\phi_j \rangle$. 
 ${\bm t}_{0n}({\bm k})$ is a  
2 by $N$ block of $\Delta{\bm H}({\bm k})$ which connects $|\phi_1\rangle$ and $|\phi_2\rangle$ with 
the $N$-fold degenerate higher energy eigenstates of 
${\bm H}({\bm k}=0)$ belonging to $\epsilon_n$, $|\phi^{(n)}_{m}\rangle$ 
($m=1,\cdots,N$); $[{\bm t}_{0n}({\bm k})]_{im} 
\equiv \langle \phi_i |\Delta {\bm H}({\bm k})|\phi^{(n)}_m \rangle$. 

A form of the 2 by 2 effective Hamiltonian thus obtained  
is constrained by the point group symmetry.  
To see this, note first that, in the 2D irreducible representation of ${\bm H}({\bm k}=0)$, 
$C_{2@1}\cdot C_{2@5}$, $C_{2@2}\cdot C_{2@4}$ and $C_{2@3}\cdot C_{2@6}$ are the 
identity operation; choosing {\it any} direction within the 2D plane will not break any of 
these point group symmetries. 
These three require that the Hamiltonian is quadratic in small ${\bm k}$. 
Namely, these symmetry operations change the sign of ${\bm k}$, 
$C_{2@1}\cdot C_{2@5} \!\ (k_x,k_y,k_z) = (k_x,-k_y,-k_z)$, 
 $C_{2@2}\cdot C_{2@4} \!\ (k_x,k_y,k_z) = (-k_x,-k_y,k_z)$ and 
 $C_{2@3}\cdot C_{2@6} \!\ (k_x,k_y,k_z) = (-k_x,k_y,-k_z)$, while they are identity 
 operation in the 2D plane. Besides, ${\bm H}^{*}({\bm k})={\bm H}(-{\bm k})$ so 
 that the effective Hamiltonian is real-valued within the second order in ${\bm k}$. 
 It takes the form of  
\begin{eqnarray}
{\bm H}^{2\times 2}_{\rm eff}({\bm k}) = d_3({\bm k}) {\bm \sigma}_3 + d_1({\bm k}) {\bm \sigma}_1 
+ d_0({\bm k}) {\bm \sigma}_0 + {\cal O}({\bm k}^3) \label{obd1}
\end{eqnarray} 
in the basis of $|\phi_1\rangle$ and $|\phi_2\rangle$; 
${\bm \sigma}_3 |\phi_{1/2}\rangle = \pm |\phi_{1/2}\rangle$. 
Here real-valued $d_{j}({\bm k})$ 
is quadratic in ${\bm k}$.

On the other hand, $C_{2@1}$, $C_{2@2}$ and $C_{2@3}$ 
are mirror operations in the 2D plane, where respective mirror axes 
form $60$ degree angle against one another. Let us call 
a mirror axis of $C_{2@1}$ to be along $|\phi_2\rangle$ and call its orthogonal vector 
as $|\phi_1\rangle$. In this basis, three operations are represented by; 
\begin{align}
C_{2@1} &=
\left(\begin{array}{cc}
-1 & 0  \\ 
0 & 1 \\
\end{array}\right), \  
C_{2@2} = \left(\begin{array}{cc} 
\frac{1}{2} & \frac{\sqrt{3}}{2}  \\ 
\frac{\sqrt{3}}{2} & -\frac{1}{2} \\
\end{array}\right),  \nn \\ 
C_{2@3} &= \left(\begin{array}{cc}
\frac{1}{2} & - \frac{\sqrt{3}}{2}  \\ 
- \frac{\sqrt{3}}{2} & -\frac{1}{2} \\
\end{array}\right). \nn
\end{align}
These mirror symmetries constrain the form of $d_{j}({\bm k})$ in eq.~(\ref{obd1}); 
\begin{align}
C_{2@1} \cdot {\bm H}^{2\times 2}_{\rm eff}(k_x,k_y,k_z) \cdot C_{2@1} 
&= {\bm H}^{2 \times 2}_{\rm eff} (-k_x,k_z,k_y), \nn \\ 
C_{2@2} \cdot {\bm H}^{2\times 2}_{\rm eff}(k_x,k_y,k_z) \cdot C_{2@2} 
&= {\bm H}^{2 \times 2}_{\rm eff} (k_y,k_x,-k_z), \nn \\ 
C_{2@3} \cdot {\bm H}^{2\times 2}_{\rm eff}(k_x,k_y,k_z) \cdot C_{2@3} 
&= {\bm H}^{2 \times 2}_{\rm eff} (k_z,-k_y,k_x). \nn 
\end{align}
These three conditions give out  
\begin{align}
{\bm H}^{2\times 2}_{\rm eff} ({\bm k}) =& 
k^2 \Big\{ d \!\ (- 2\hat{k}^2_x + \hat{k}^2_y + \hat{k}^2_z) {\bm \sigma}_3 \nn \\
& - \sqrt{3} d \!\ (\hat{k}^2_y - \hat{k}^2_z) 
{\bm \sigma}_1 + A {\bm \sigma}_0 \Big\} + {\cal O}( k^3), \label{eq1}
\end{align} 
with $k \equiv |{\bm k}|$ and $\hat{k} \equiv {\bm k}/k$. Here $d$ and 
$A (>2|d|)$ depend on a microscopic parameter. They satisfy  
$A-2|d|>0$ in the Z$_6$ phase region, where the lowest eigenmodes 
of ${\bm H}({\bm k})$ are at ${\bm k}=0$.

\begin{figure}[t]
\begin{center}
\includegraphics[width=85mm]{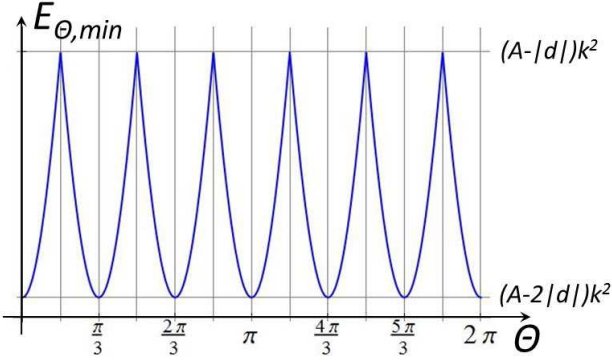}
 \caption{The lowest magnetic energy of finite-momentum spin configuration around  
 $|\theta\rangle \equiv \cos\theta \!\ |\phi_1\rangle + \sin\theta \!\ |\phi_2\rangle$ 
 is shown as a function of $\theta$ for $d>0$. The vertical axis 
 denotes $E_{\theta,{\rm min}}
  \equiv \min_{\hat{\bm k}} E_{\theta,{\bm k}}$ with 
  $E_{\theta,{\bm k}}\equiv \langle \theta | 
  {\bm H}^{2\times 2}_{\rm eff}({\bm k}) | \theta \rangle$.}
 \label{fig6}
 \end{center}
\end{figure}

A magnetic energy for a linear combination of $|\phi_1\rangle$ and $|\phi_2\rangle$ with finite (but small)  
momenta can be calculated from ${\bm H}^{2\times 2}_{\rm eff}({\bm k})$ as 
$E_{\theta,{\bm k}} \equiv \langle \theta |{\bm H}^{2\times 2}_{\rm eff}({\bm k}) 
|\theta \rangle$ with $|\theta\rangle \equiv \cos\theta |\phi_1 \rangle + \sin\theta |\phi_2 \rangle$. 
The energy for such a spin configuration can be further minimized with respect to $\hat{\bm k}$
. The minimized energy has 6 minima at $\theta = 0,\pm\frac{\pi}{3}, \pm\frac{2\pi}{3}, \pi$ 
for $d>0$, while at $\theta = \pm\frac{\pi}{6},\pm\frac{\pi}{2}, \pm\frac{5\pi}{6}$ for $d<0$ 
(Fig.~\ref{fig6}). The optimal directions  
of $\hat{\bm k}$ for the former 6 minima are along $k_z$, $k_x(k_y)$, $k_y(k_x)$, $k_z$ axes 
respectively. Due to the entropy effect, 
either one of these two types of six minima is selected as the ordering direction at finite temperature. 

To see this entropy effect, one can also calculate the free energy which includes spatial 
fluctuation around $|\theta\rangle$ within the second order in small ${\bm k}$;
\begin{align}
F(\theta) = -\frac{1}{\beta} \ln \frac{1}{V} \sum_{\bm k} e^{-\beta E_{\theta,{\bm k}}} = \frac{1}{2\beta} \ln \Big[\prod_{j=x,y,z} \big(4\pi \beta B_j\big)\Big] \nn
\end{align} 
with $B_{x} \equiv A-2d \cos2\theta$, $B_y \equiv A -2d \cos(2\theta -\frac{2\pi}{3})$ and 
$B_z \equiv A -2d \cos(2\theta +\frac{2\pi}{3})$. At $T\ne 0$, $F(\theta)$ has indeed six minima 
at $\theta = 0,\pm \frac{\pi}{3}, \pm \frac{2\pi}{3}, \pi$ 
for  $d>0$, while at $\theta = \pm \frac{\pi}{6},\pm \frac{\pi}{2}, \pm\frac{5\pi}{6}$ for $d<0$. 
Consistently, the MC simulation found that the system at the 
lower temperature prefers either one of these two types of six minima (Fig.~\ref{fig5}(b)).

When $|\phi_1\rangle$ (or its 5 counterparts; $\theta=0,\pm\frac{\pi}{3},\pm \frac{2\pi}{3}$) 
is selected by the classical order by disorder, 
the spin configuration breaks all $C_{2@j}$ ($j=1,2,3$) but is invariant 
under $C_{2@1}\cdot T$; the generator of the reduced magnetic point 
group is $\{C_{2@1}\cdot C_{2@5}, C_{2@2}\cdot C_{2@4}, C_{2@3}\cdot C_{2@6},  
C_{2@1}\cdot T\}$. The first three in the set change the sign of spin
moments, i.e. $C_{2@1}\cdot C_{2@5} \!\ (S_x,S_y,S_z) = (S_x,-S_y,-S_z)$, 
$C_{2@2}\cdot C_{2@4} \!\ (S_x,S_y,S_z) = (-S_x,-S_y,S_z)$ and $C_{2@3}\cdot C_{2@6} \!\ (S_x,S_y,S_z) = 
(-S_x,S_y,-S_z)$. Thus, the phase is an antiferromagnetic (AF) phase without any finite 
off-diagonal elements of magnetic susceptibility tensor; 
$\chi_{xy} =\chi_{yz} = \chi_{zx} =0$. The last one in the set, $C_{2@1}\cdot T$,  
connects $y$ and $z$ while it connects $x$ with neither $y$ nor $z$; the phase shows 
spin anisotropy in diagonal susceptibility, 
\begin{eqnarray}
\chi_{xx} \ne \chi_{yy} =\chi_{zz}. \label{z6a}
\end{eqnarray}
When $|\phi_2\rangle$ (or its 5 
counterparts; $\theta=\pm\frac{\pi}{2},\pm\frac{\pi}{6},\pm \frac{5\pi}{6}$)  
is chosen, the configuration breaks 
$C_{2@2}$ and $C_{2@3}$ , but is invariant under $C_{2@1}$;    
$\{C_{2@1}\cdot C_{2@5}, C_{2@2}\cdot C_{2@4}, C_{2@3}\cdot C_{2@6},  
C_{2@1}\}$. Accordingly, the phase is an AF phase without off-diagonal 
susceptibility tensor and with $
\chi_{xx} \ne \chi_{yy} =\chi_{zz}$. 
In the presence of large 
lattice-spin coupling, both of these two Z$_6$ SB  phases 
are accompanied with an uniaxial lattice distortion differentiating $x$ coordinate 
from $y$ and $z$ coordinates. 
      
\section{emergent $U(1)$ symmetry around the critical point}
The U(1) to Z$_6$ crossover behavior around $T_*$ 
in the Z$_6$ SB magnetic phase 
can be crudely captured by 
the six states ferromagnetic Potts model in the 3D 
lattice. Low-$T$ physics of the Potts model are controlled by 3D XY fixed 
point, $T=0$ Nambu-Goldstone (NG) fixed point and $T=0$ fixed point with 
large Z$_6$ term.~\cite{1984,oshikawa} The critical point ($T=T_c$)  belongs 
to the XY fixed point, below which  
all the renormalization group (RG) flow goes to the large Z$_6$  
fixed point.~\cite{1984} 
The U(1) to Z$_6$ crossover below $T_c$ stems from the dangerously 
irrelevant behavior of the Z$_6$ anisotropy term.~\cite{1984,oshikawa} 
The anisotropy term is renormalized to smaller value around the XY fixed point, 
while it blows up into larger value around the NG fixed point. Due to this behavior, 
a smaller system near $T_c$ behaves as if it has much reduced 
Z$_6$ anisotropy, while a larger system far from 
$T_c$ behaves as a system with enhanced anisotropy.

\begin{figure}[t]
\begin{center}
\includegraphics[width=80mm]{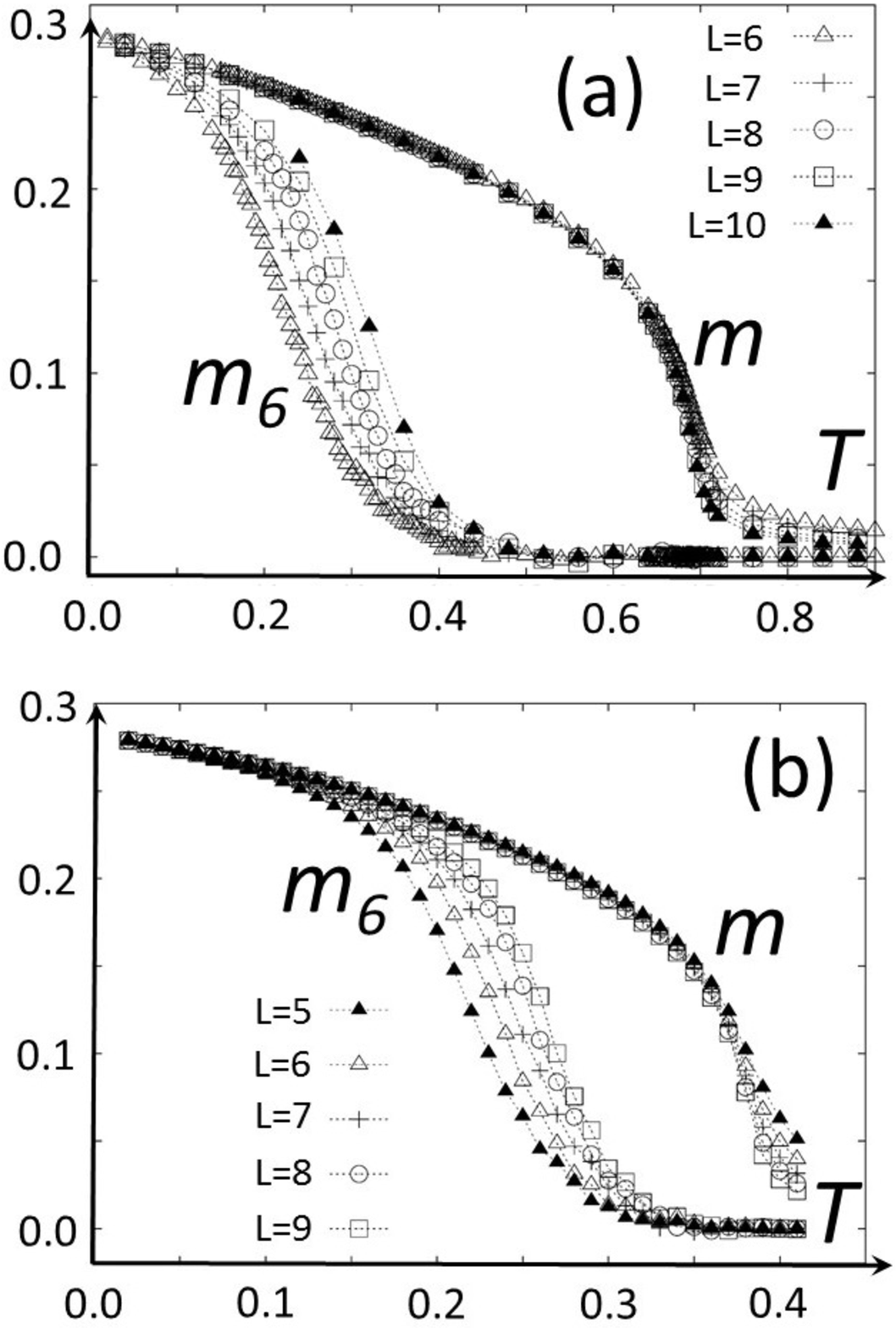}
 \caption{Temperature dependence of XY order parameter $m$ and Z$_6$ SB order parameter ($m_6$) 
 at different parameter points; (a) $(J,D,G)=(0.50,-0.22,-0.84)$ (b) $(J,D,G)=(0.69,-0.53,-0.50)$.}
 \label{fig7}
 \end{center}
\end{figure}

Fig.~\ref{fig7} shows a set of temperature dependencies of the XY order parameter 
and Z$_6$ order parameter;  
\begin{align} 
m^2 &\equiv m^2_1+m^2_2 \nn \\
m_6 &\equiv [(m_1+im_2)^6 + (m_{1}-im_{2})^6]/(2m^5), \nn 
\end{align}
calculated for different system 
sizes ($L=5,6,7,8,9,(10)$) at two different parameter points in the Z$_6$ SB phase 
region; $(J,D,G)=(0.50,-0.22,-0.84)$ and $(0.69,-0.53,-0.50)$. The $T$-dependences 
of $m$ and $m_6$ clearly show that, in finite system sizes, the XY order parameter and 
the Z$_6$ order parameter start to take finite values at different temperatures.  
Note also that the former parameter point is proximate 
to the SU(2) symmetry point  $(J,D,G)=(0,0,-1)$; the SU(2) point is approximately 
an effective spin model of Na-438, when the atomic spin-orbit interaction is larger 
than the non-cubic crystal field splitting energy and when the exchange path is 
mainly mediated by the oxygen ions~\cite{cb}. Meanwhile, 
the latter parameter point is far from  any 
high symmetric parameter points, playing the role of a good 
reference parameter point (see below).
 
\begin{figure}[t]
   \centering
   \includegraphics[width=80mm]{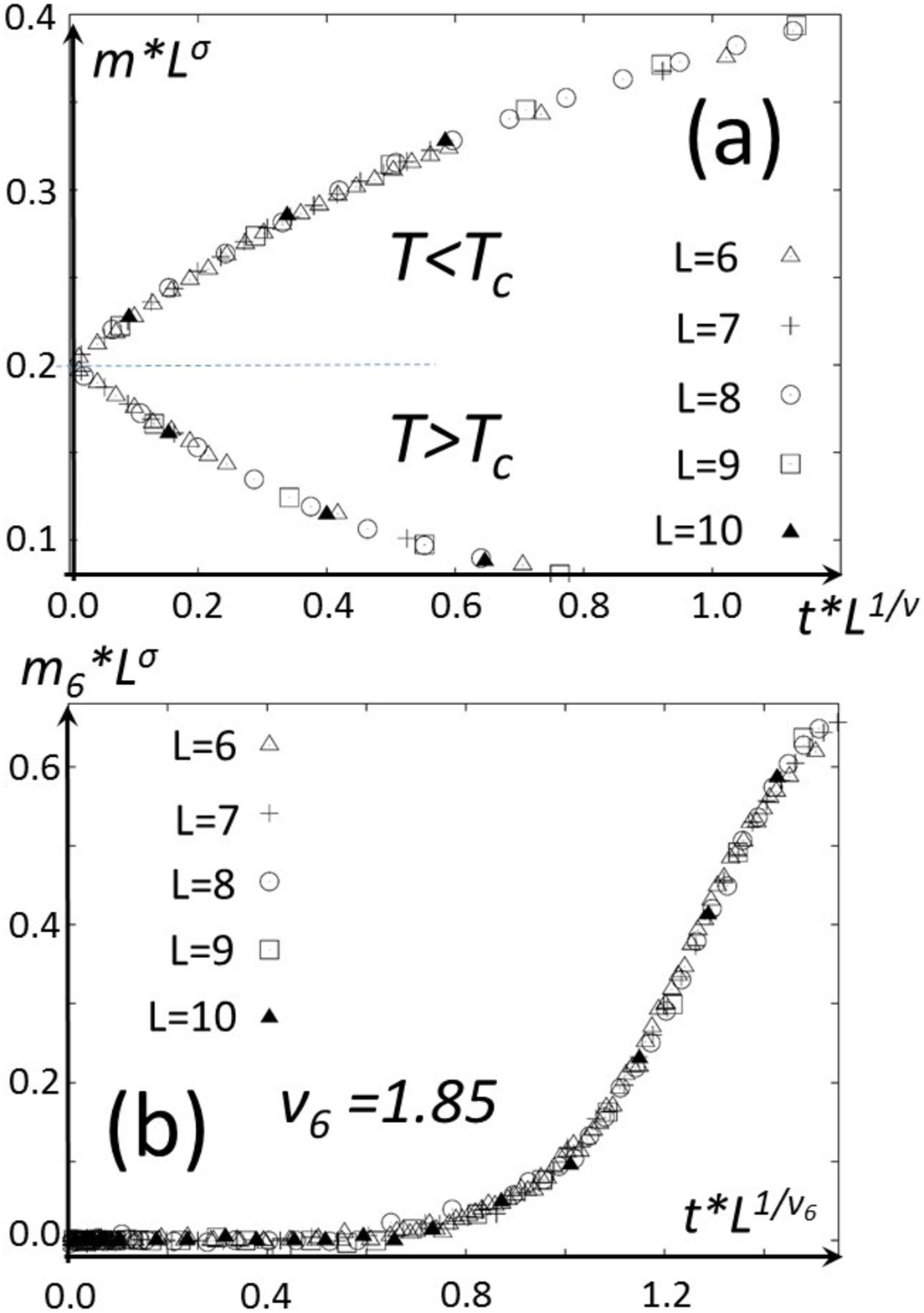} 
  \caption{Single-parameter scalings for XY order parameter $m$ (a) and Z$_6$ SB order 
  parameter $m_6$ (b) at $(J,D,G)=(0.50,-0.22,-0.84)$. }
  \label{fig8} 
\end{figure}

\begin{figure}[t]
\begin{center}
\includegraphics[width=80mm]{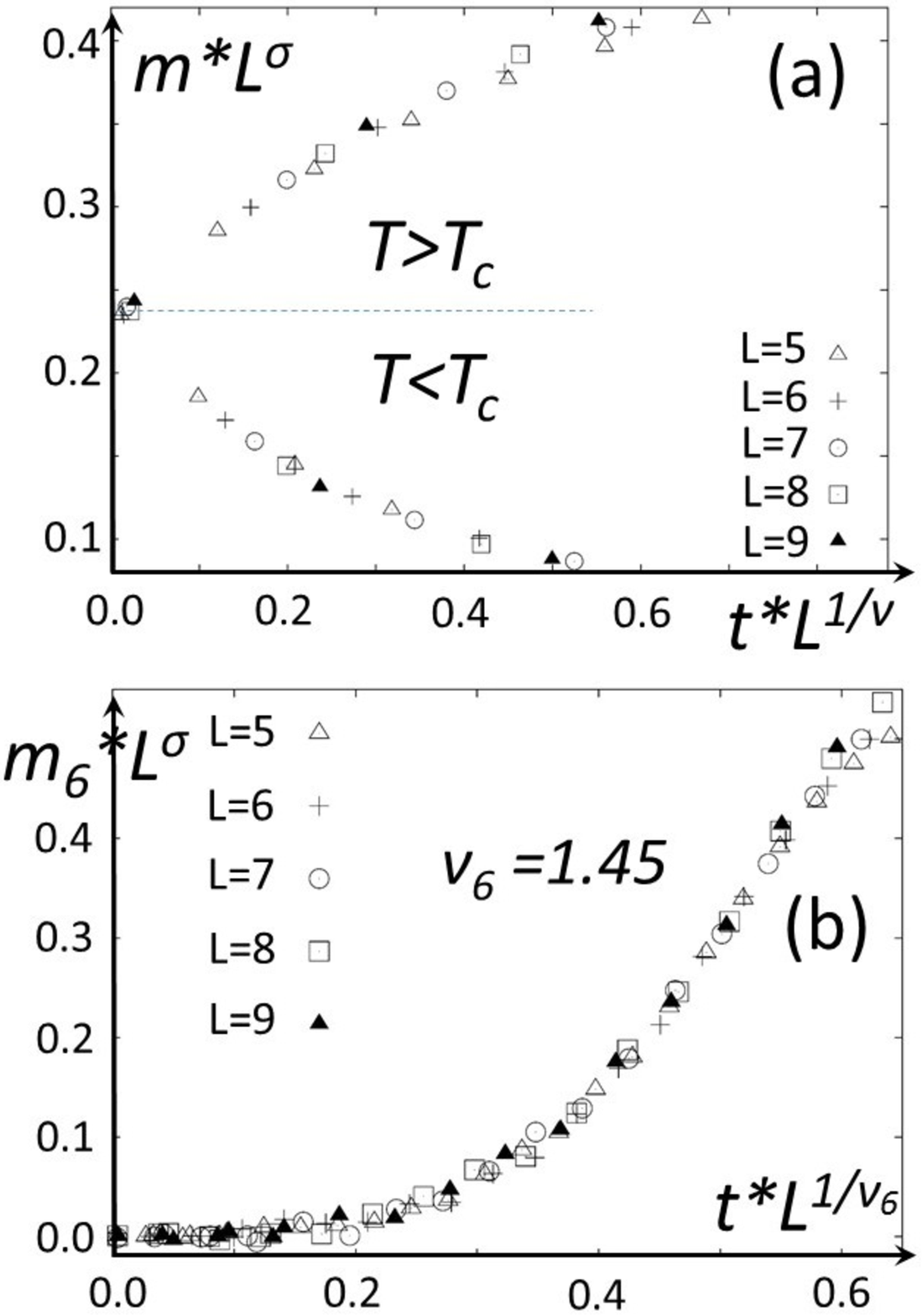}
 \caption{Single-parameter scalings for the XY order parameter $m$ (a) and Z$_6$ order parameter $m_6$ (b) 
 at $(J,D,G)=(0.69,-0.53,-0.50)$.}
 \label{fig9}
 \end{center}
\end{figure}

The crossover system size and temperature can be evaluated from the FNS 
analysis on the XY order parameter $m$ and Z$_6$ order 
parameter $m_6$. The scaling argument~\cite{oshikawa,lou,okubo} suggests 
that these two follow single-parameter scalings; 
\begin{align}
m &= L^{-\sigma} f(t L^{1/\nu}), \label{sps1} \\
m_{6}& = L^{-\sigma} f_6(t L^{1/\nu_6}), \label{sps2}
\end{align}  
with $\sigma=\beta/\nu$ and $t \equiv |T-T_c|/T_c$. $\beta$ and $\nu$ denote  
the critical exponents of the 3D XY universality class. 
Fig.~\ref{fig8} and \ref{fig9} show respective one-parameter scaling fittings 
at the two parameter points. 
To obtain them, we fixed the critical exponents $\beta$ and $\nu$ to be those  
for the 3D XY universality class ($\nu=0.672$ and $\beta=0.348$)\cite{xy}, while 
fine-tuned the critical temperature $T_c$ 
such that all the data points for $m$ fall into one scaling function $f(x)$. 
With $T_c$ thus obtained, we further fine-tuned the crossover exponent for $m_6$, $\nu_6$, 
such that all the data points for $m_6$ fall into the one-parameter 
scaling function $f_6(x)$. The two fittings demonstrate that all the numerical data points 
for $m$ and $m_6$ taken from different system size ($L=5,6,\cdots,10$) 
fall into respective one-parameter scaling functions.  
The optimal $\nu_6$ is evaluated at two different parameter points, 
$(J,D,G)=(0.69,-0.53,-0.50)$, $(0.50,-0.22,-0.84)$, as
1.45 $\pm 0.05$  
and 1.85 $\pm 0.05$ respectively.  
The crossover system size $\Lambda_*$ and temperature $T_*$ can be obtained 
by equating the argument of $f_6(\cdots)$ with  
unit; $t\Lambda^{1/\nu_6}_{*}=1$ or $t_{*}\Lambda^{1/\nu_6}=1$. With $\xi \sim t^{-\nu}$, we 
obtain $\Lambda_* \sim \xi^{\nu_6/\nu}$ and $\Delta T_{*} \sim \Lambda^{-1/\nu_6}$ respectively.  

The former value of the crossover exponent $\nu_6$ (1.45 $\pm 0.05$) is  
consistent with previous estimation in the Potts model~\cite{lou,okubo},  
while the latter value (1.85 $\pm 0.05$)  is relatively larger. 
The discrepancy stems from the presence of a high symmetric point 
at $(J,D,G)=(0,0,-1)$ near the latter parameter point. 
The symmetric point has a global SU(2) symmetry 
toward which the Z$_6$ anisotropy 
diminishes. Such a symmetric point plays the role of another crossover fixed 
point for RG flow, changing the crossover exponent from 
that of the simple Z$_6$ Potts model. 

\section{Effects of Quantum Fluctuation} 

In the quantum $J=1/2$ pseudo-spin case, the techniques used in this 
paper may not be able to capture physics at or near highly frustrated 
parameter points, where all magnetic ordering temperatures reduce to 
zero due to strong spin frustration. Such frustrated points include 
the AF Heisenberg point $(J,D,G)=(1,0,0)$ 
and Kitaev point $(J,D,G)=(\frac{1}{\sqrt{2}},0,\frac{1}{\sqrt{2}})$. 
On the one hand, a comparison between preceding classical calculations~\cite{price1,price2} 
and quantum calculations~\cite{a-li-213-b,kitaev} on the honeycomb iridate spin 
model suggests that, even for the quantum $J=1/2$ case, physics discussed 
in this paper may hold true in those parameter regions 
with higher magnetic ordering temperatures. Such parameter 
regions cover most of the physically relevant parameter regions ($3J-G>0$), including 
one of the `candidate' parameter points, i.e. the SU(2) point 
$(J,D,G) = (0,0,-1)$. In these regions, the quantum 
fluctuation changes the results in a quantitative level as discussed 
below. 
 
Firstly, the Z$_2$-Z$_6$ phase boundaries in Fig.~\ref{fig3} at $T=0$  
are expected to move into the Z$_2$ phase side due to the quantum fluctuation. 
This is because the finite-$T$ first order phase boundary 
between Z$_2$ and Z$_6$ phases is slanted in a way that 
the Z$_6$ phases are more stabilized against the Z$_2$ phase by the thermal effect 
(Fig.~{\ref{fig4}}(b,c)).  
With the Clausius-Clapeyron relation, this indicates that, around the phase boundary, 
the Z$_6$ SB states have more low energy classical spin configurations 
`proximate' to themselves than the competing Z$_2$ SB state has. 
Here we define a given classical spin configuration 
``$x$'' to be more `proximate' to ``$y$'' than to ``$z$'' when $x\cdot y > x\cdot z$, 
where ``$\cdot $'' denotes the inner product 
between two spin configurations with respect to spin 
index and site index. 

Crudely speaking, the quantum zero-point 
energy~\cite{HP} can be regarded as the second-order 
energy correction to the classical ground state, which comes from the 
virtual hopping processes between the classical ground state $|0\rangle $ 
and another classical spin states with higher classical energies $|n\rangle $;  
\begin{eqnarray}
E_{{\rm SW}} = - \sum_n 
\frac{|\langle 0|H^{\prime}|n\rangle |^2}{E_n-E_0} + {\cal O}({H^{\prime}}^4). \label{quantum}
\end{eqnarray}
Here $H^{\prime}$ stands for the quantum fluctuation term, 
comprising `off-diagonal' terms such as $a^{\dagger}_i a^{\dagger}_j$, 
$a_i a_j$, and $a^{\dagger}_i a_j$ with $i \ne j$, while $a^{\dagger}_i$ being 
the Holstein-Primakoff boson (creation) operator at the $i$-th site. The zero-th order 
part $H_0$ comprises `diagonal' terms such as $a^{\dagger}_i a_i$, giving 
the classical energies to each classical spin states e.g. 
$H_0|n\rangle  = E_n |n\rangle$. 

The indication of the finite-$T$ phase boundary between Z$_2$ and Z$_6$ 
phases in combination with eq.~(\ref{quantum}) suggests that, 
near the Z$_2$-Z$_6$ phase boundaries at $T=0$ in Fig.~\ref{fig4}(b,c), 
the quantum zero-point energy of the Z$_6$ states will 
be larger than that of the Z$_2$ state. Namely, the Z$_6$ states have 
possibly more low-energy classical spin configurations ($|n\rangle$) connecting 
with Z$_6$ states by the local perturbation $H^{\prime}$ than the 
Z$_2$ state does. As a result, the Z$_2$-Z$_6$ classical phase boundaries  
in Fig.~\ref{fig4}(b,c) are expected to move into the Z$_2$ phase side, when the 
quantum fluctuation is included perturbatively. In fact, one can 
find in a recent literature~\cite{xyPY} a model calculation on a 
different spin system, whose observation agrees with the thoughts 
given above. One may also notice from Fig.~\ref{fig4}(a) that the 
finite-$T$ phase boundary between Z$_2$ and Z$_6$ phases 
is constrained in the $D=0$ plane. This exceptional feature 
is, however, due to the additional symmetry at $D=0$ (Appendix A). 

Another possible quantitative change could be found in the finite-size crossover 
phenomena in the Z$_6$ phase, which may be effectively characterized by 
the $(d+1)$-dimensional Z$_6$ ferromagnetic Potts model. In the effective 
model, the quantum effect is taken into account as an addition of the imaginary 
time dimension (+1) to the spatial dimension ($d=3$). The 4D Z$_6$ Potts model 
exhibits the same kind of finite-size crossover phenomena with 
different crossover exponent~\cite{okubo}. 
Moreover, in the (3+1)D model, finite temperature 
leads to a non-trivial `finite-size' effect along the imaginary time direction, 
in the same way as the finite system size does along the spatial direction. 
Thus, one may even expect that the Z$_6$ ordered phase  
accommodates two distinct crossover temperatures below a finite critical 
ordering temperature $T_c$; one is associated with the spatial fluctuation 
of spins and the other with their temporal fluctuation. 

\section{Discussion}
In this paper, we obtained a comprehensive classical 
magnetic phase diagram for the 
hyperkagome iridate. We clarified the origin of the Z$_6$ anisotropy 
and finite-$T$ ordering nature of the Z$_{6}$ phases. 
Our finite-$T$ classical phase diagram suggests that the 
Z$_6$ phases could be further stabilized against the competing Z$_2$ phase 
by the quantum order by disorder. 

Based on the finite-$T$ crossover behavior in the Z$_6$ SB phase, let us 
finally introduce a possible phenomenology of  
Na-438 powder samples.  Firstly we assume that polycrystalline 
grain size is as small as 1$\mu$m $\sim$ 0.1$\mu$m and we regard that a 
broad peak in specific heat observed around $T\simeq 35K$ in experiments~\cite{ot,sg,dally} 
corresponds to the onset temperature of the Z$_6$ SB phase ($T_c$).  
The FNS argument above claims that, below but near $T_c$, a system smaller than 
the crossover system size $\Lambda_* = \xi^{\nu_6/\nu}$   
behaves as if it has no Z$_6$ anisotropy term. 
Spins in such a small grain collectively develop a finite 
amplitude of  $\Phi \equiv \phi_1+i\phi_2$ below $T_c$, while the phase of 
$\Phi$ still strongly fluctuates and so does that of 
individual spins. Thus, any local spin moment seen 
by probe spins can be averaged to be zero 
in the intermediate temperature regime $T_*<T<T_c$.  
For $\xi=3,5,7,10$ cubic unit cells, the crossover system size  
is evaluated to be $\Lambda_*=20, 90, 230, 630$ cubic unit cells 
for $\nu_6=1.85$  ($0.017$, $0.08$, $0.20$, $0.56$ $\mu$m 
with a=8.95 $\AA$). On further lowering temperature, the correlation length $\xi$ 
becomes shorter and so does $\Lambda_*$. When $\Lambda_*$ exceeds the  
grain size on lowering temperature ($T<T_*$), the Z$_6$ anisotropy becomes prominent 
and the phase of $\Phi$ starts to be locked into the six minima. As explained above, 
this locking breaks the 
point group symmetry, giving rise to the uniaxial lattice distortion via spin-lattice 
coupling. Since neighboring grains in polycrystalline sample are expected to be 
randomly oriented against one another, the locking or its onset around $T_{*}$ 
with the uniaxial distortion will conflict with the grain structure. This may  
result in a `configurationally degenerate phase with 
fluctuating order'  as suggested in the 
experiment~\cite{dally}.   
  
The author would like to thank 
Gang Chen, Kenji Harada and Xuerong Liu for fruitful discussions.  
This work was financially supported by NBRP of China (2015CB921104).
    
\appendix

\section{Finite-$T$ phase diagram at $D=0$}
\begin{figure}[t]
\begin{center}
\includegraphics[width=85mm]{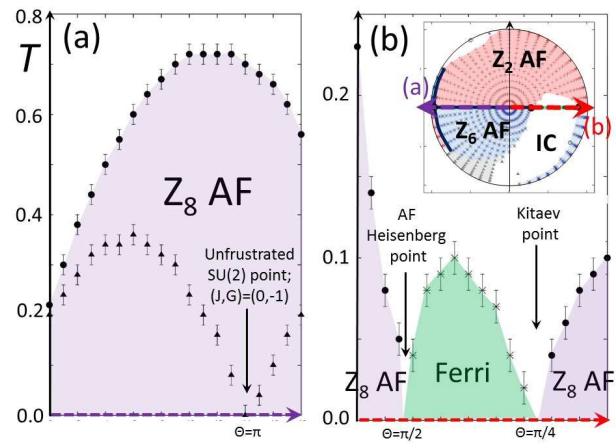}
 \caption{Finite-$T$ phase diagrams at $D=0$ and $3J-G>0$. (a) $\frac{\pi}{2} + \varphi<\theta<\pi+\varphi$ 
 (b) $\varphi <\theta< \frac{\pi}{2} +\varphi$ with $(J,G)=(\sin\theta,\cos\theta)$ and 
 $\tan\varphi=\frac{1}{3}$. The critical temperatures of the Z$_8$ AF phase and ferrimagnetic 
 phase are determined from specific heat peak (black filled circle points and black double crossed 
 points). 
 The crossover temperature within the Z$_8$ AF phase is determined 
from temperature dependence of sublattice magnetization (black upper triangle points).}
 \label{fig10}
 \end{center}
\end{figure}

Finite-$T$ phase diagrams at $D=0$ are shown in Fig.~\ref{fig10}. A group of 
Hamiltonians at $D=0$ is symmetric under a `Klein' transformation $G$~\cite{kimchi}; 
$H_{\theta} = G\cdot H_{\frac{\pi}{2}-\theta}\cdot G$ 
with $(J,G) \equiv (\sin\theta,\cos\theta)$. This connects antiferromagnetic 
(AF) and ferromagnetic (F) Heisenberg point $(J,G)=(\pm1,0)$ with two other 
symmetric points with global SU(2) symmetries $(J,G)=(0,\pm 1)$, 
while leaves intact AF Kitaev point $(J,G)=(\frac{1}{\sqrt{2}},\frac{1}{\sqrt{2}})$ 
and F Kitaev point $(J,G)=-(\frac{1}{\sqrt{2}},\frac{1}{\sqrt{2}})$. Consistently,  
the finite-$T$ phase diagram at $D=0$ is symmetric 
under this transformation, 
where the ferrimagnetic phase $(\frac{\pi}{4}<\theta<\frac{\pi}{2})$ is 
transformed by $G$ to the Z$_8$ antiferromagnetic phase ($0<\theta<\frac{\pi}{4}$),   
the ferromagnetic phase $(-\frac{3\pi}{4}<\theta<0)$ 
to the $Z_8$ antiferromagnetic phase $(\frac{\pi}{2}<\theta<\frac{5\pi}{4})$. The ordering 
temperatures of respective phases vanish at two AF SU(2) points ($\theta=0,\frac{\pi}{2}$) 
and two Kitaev points ($\theta=\frac{\pi}{4},\frac{5\pi}{4}$). On the one 
hand, two F SU(2) points ($\theta=\pi,\frac{3\pi}{2}$)  are 
unfrustrated points, where magnetic orderings show the maximum transition 
temperature. No O(3) to Z$_8$ crossover is observed within  
$0<\theta<\frac{\pi}{2}$, while ,for $\frac{\pi}{2}<\theta<2\pi$, O(3) to Z$_8$ crossover   
are observed below ordering temperatures  for a finite-size system. 
The crossover region becomes widest at the unfrustrated F SU(2) point.

\begin{figure}[htbp]
\begin{center}
\includegraphics[width=80mm]{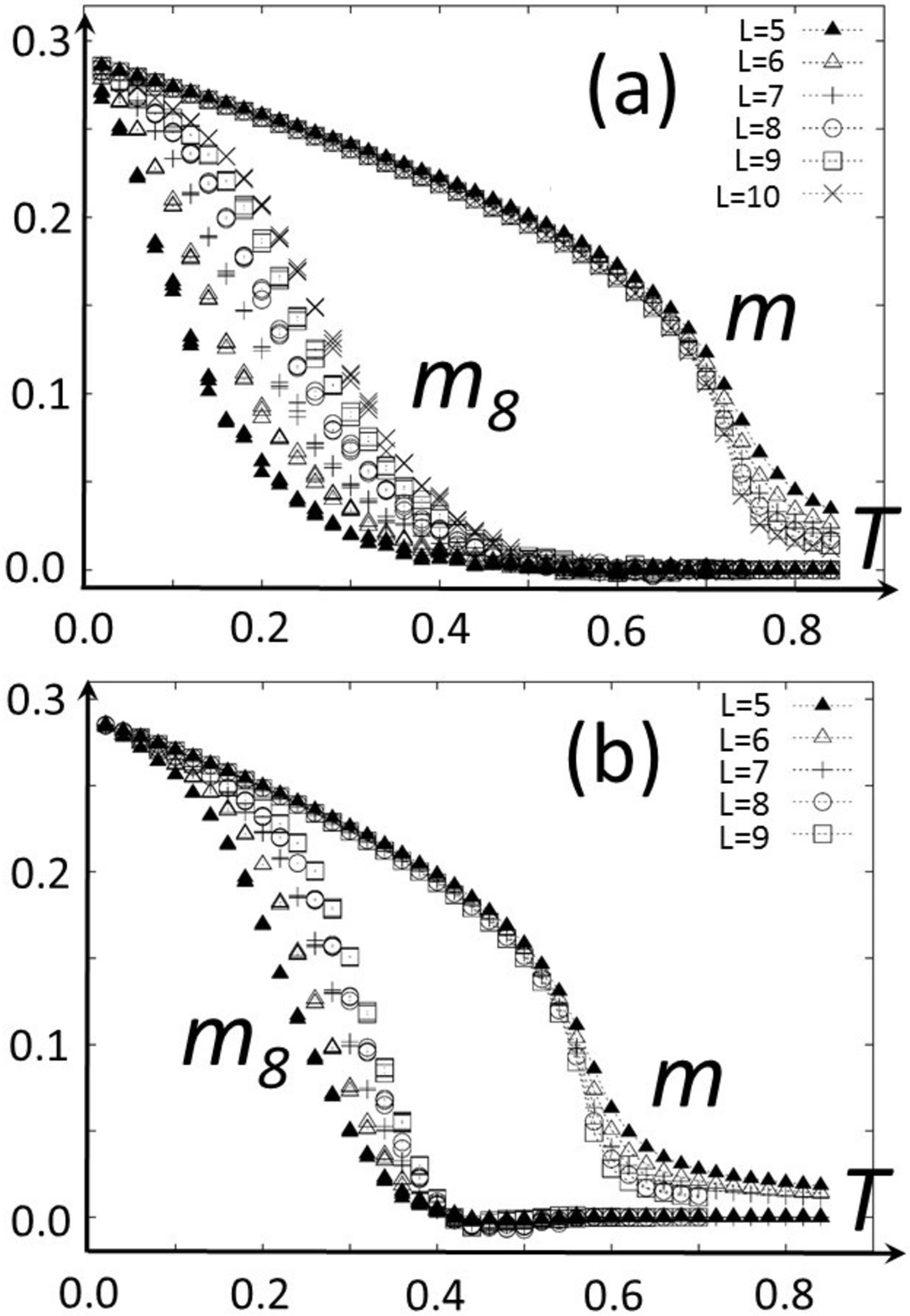}
 \caption{Temperature dependence of O(3) order parameter $m$ and Z$_8$ SB order parameter $m_8$  
 at different parameter points; (a) $(J,D,G)=(0.11,0.0,-0.99)$ (b) $(J,D,G)=(0.31,0.0,-0.95)$.}
 \label{fig11}
 \end{center}
\end{figure}
\begin{figure}[htbp]
\begin{center}
\includegraphics[width=80mm]{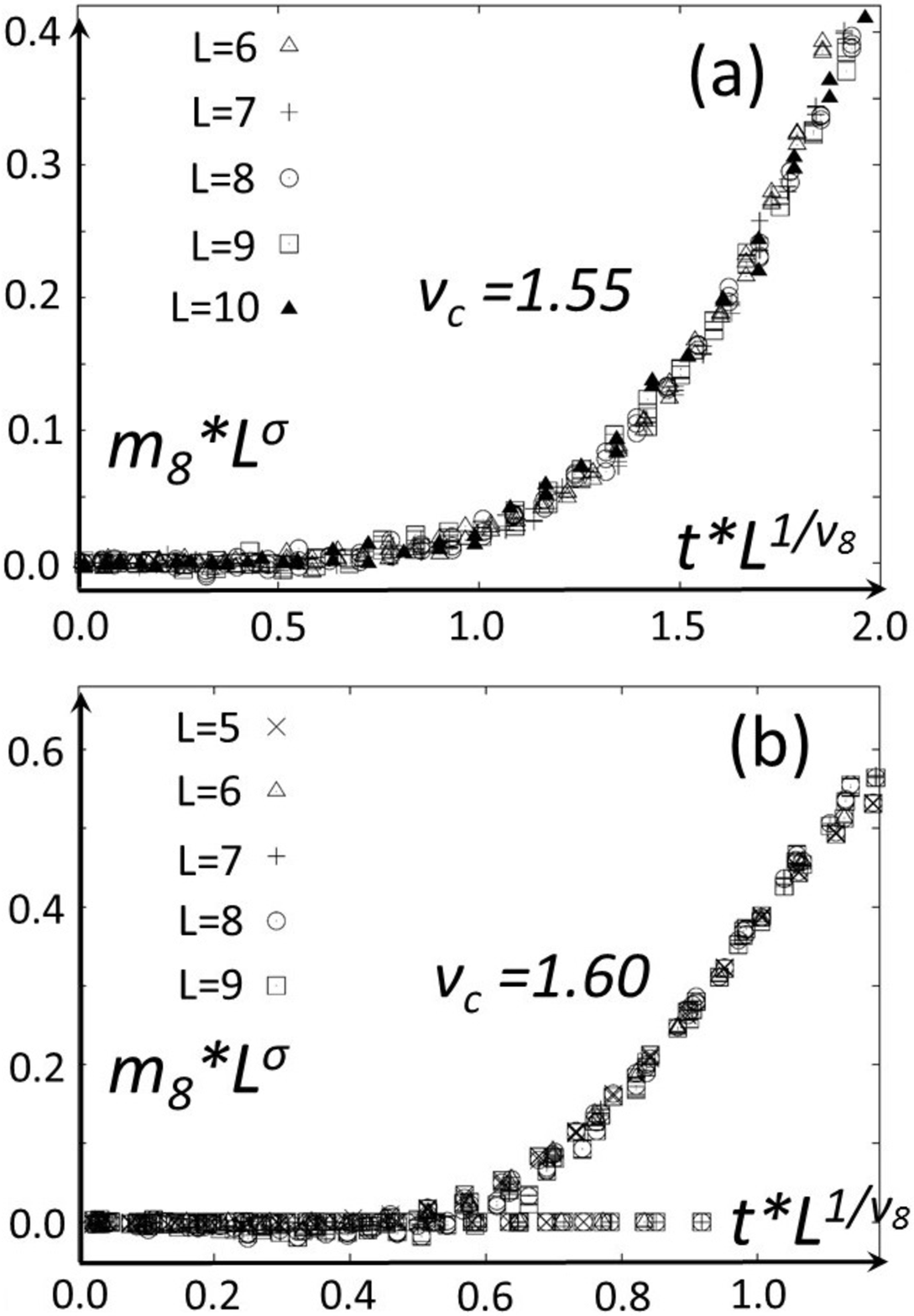}
 \caption{Single-parameter scalings for the Z$_8$ SB order parameter at $D=0$
 with different parameter points; (a) $(J,G)=(0.11,-0.99)$ (b) $(J,G)=(0.31,-0.95)$.}
 \label{fig12}
 \end{center}
\end{figure}

\section{finite size scaling analysis in Z$_8$ SB phase} 

The MC simulation shows that the phase boundary between the Z$_2$ and Z$_6$ SB phase is 
always of the first order (Appendix C), indicating the existence of a bicritical point at finite temperature. 
Below the bicritical point, the system exhibits Z$_8$ SB magnetic phase with 
O(3)-type spin fluctuation. This situation can be clearly seen at the phase boundary of $D=0$ 
by the LT analysis. In the phase boundary between 
Z$_2$ and Z$_6$ phases, ${\bm H}({\bm k})$ has triple degeneracy in its lowest eigenmodes at 
${\bm k}=0$; two are the doubly degenerate lowest eigenmodes from the Z$_6$ phase region 
($|\phi_1\rangle$ and $|\phi_2\rangle$) and one is the lowest eigenmode from the Z$_2$ 
phase region ($|\phi_3\rangle$). These three form a O(3) 
sphere in the 36 dimensional space. On the sphere, the fixed norm condition is satisfied 
only along 8 high-symmetry directions; $\pm (1,1,1)$, $\pm (1,1,-1)$, $\pm (1,-1,1)$, $\pm (-1,1,1)$ 
($(1,1,1)$ means $|\phi_1\rangle + |\phi_2\rangle + |\phi_3\rangle$), while otherwise not 
in general.  These directions become strong easy-axis directions, giving rise to 
the cubic anisotropy in the O(3) vector model.          

The ordering nature of the 3D O($n$) vector model with the cubic anisotropy is determined 
by either 3D Heisenberg fixed point ($n<n_c$) or cubic fixed point $(n>n_c)$. 
Preceding studies including the fifth order $4$-$\epsilon$ expansion 
conclude that $n_c$ is smaller than 3 ($n_c=2.89$).~\cite{cubic} 
Being consistent with this, the MC simulation on $G>0$ and $D=0$ region 
does not observe any O(3) to Z$_8$ crossover behavior (see Fig.~\ref{fig10}(b)); 
Below a critical temperature (determined by the specific heat peak), 
the Z$_8$ anisotropy always becomes prominent in the O(3) sphere  
even for the smallest simulated system size ($L=6$). 

On the one hand, we also observe an unexpected O(3) to Z$_{8}$ crossover 
region below the bicritical point at $G<0$ and $D=0$ (Fig.~\ref{fig10}(a)). 
Fig~\ref{fig11} shows a set of temperature dependences of the O(3) order parameter $m$ 
and Z$_8$ order parameters $m_{c,\mu}$ ($\mu=1,2,3$) 
calculated for different system sizes and at two different parameter points in 
the Z$_8$ SB phase region at $G<0$ and $D=0$; 
\begin{align}
m &\equiv \sqrt{m^2_1 + m^2_2 + m^3_2} \nn \\
m_{c,\mu} &\equiv \epsilon_{\mu\nu\rho}\frac{(m_\nu+im_\rho)^4 + (m_{\nu}-im_{\rho})^4}{2m^3}. \nn
\end{align} 
Three Z$_8$ SB order parameters ($\mu=1,2,3$) show the 
same temperature dependence.  
Fig.~\ref{fig12} shows one-parameter scaling forms for the Z$_8$ SB order parameters; 
$m_{c,\mu} = L^{-\sigma} g_c(t L^{1/\nu_c})$. 
To obtain them, we fixed $\nu$ and $\beta$ to be those of 
3D cubic fixed point ($\nu = 0.704$ and $\beta= 0.362$)~\cite{cubic}, 
while we fine-tuned $T_c$ and $\nu_{c}$ to fit 
data points for $m$ and $m_{c,\mu}$ ($\mu=1,2,3$) into $g(x)$ (not shown) 
and $g_c(x)$ (Fig.~\ref{fig12}) respectively. The optimal $\nu_c$ thus determined 
is $1.55 \pm 0.05$ at the two different parameter points.   

The O(3) to Z$_8$ crossover at $G<0$ and $D=0$ is 
apparently counterintuitive from the viewpoint of the previous studies on the 
$O(n)$ vector model with the cubic anisotropy.~\cite{cubic} 
A simple explanation for this is that the cubic anisotropy at the microscopic Hamiltonian 
level is much smaller than $T_c$ and because the largest simulated system size ($L=10$) is 
still too small that renormalized (thus enhanced) cubic term is tiny compared to $T_c$.    
In fact, the cubic term vanishes completely at the unfrustrated SU(2) point; 
$(J,D,G)=(0,0,-1)$, while 
$T_c$ takes the largest value at this SU(2) point (see $\theta=\pi$ in Fig.~\ref{fig10}(a)). 
Besides, a positive scaling dimension of the cubic term around the 3D Heisenberg fixed point was 
suggested to be very small.~\cite{cubic}

\begin{figure}[t]
\begin{center}
\includegraphics[width=85mm]{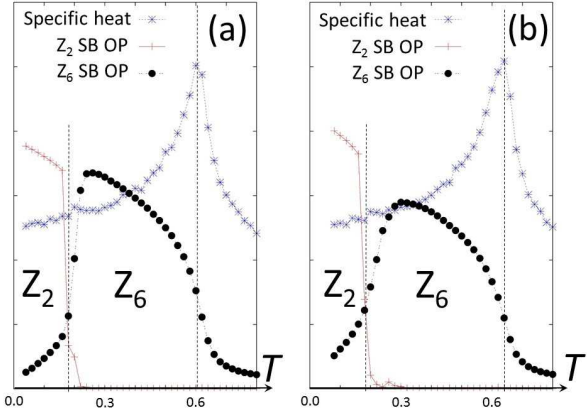}
 \caption{Temperature dependencies of Z$_2$ SB order parameter (red cross points), 
 XY order parameter (black filled circle points)  and specific heat (blue double cross points) 
 around the phase boundary between Z$_2$ SB phase and Z$_6$ SB phase; 
 (a) $(J,D,G)=(0.05,-0.36,-0.93)$ (b) $(J,D,G)=(-0.05,0.25,-0.97)$}
 \label{fig13}
 \end{center}
\end{figure}

\section{MC evidence for the first order phase transition between Z$_2$ and Z$_6$ SB phases}
Fig.~\ref{fig13} shows a set of temperature dependence of the Z$_2$ SB order parameter and that of XY order 
parameter (order parameter for the Z$_6$ SB phase) around finite-$T$ phase boundaries between 
Z$_2$ phase and Z$_6$ phase. The results show discontinuous changes of these two 
order parameters at the phase boundaries, indicating that the transition is of the first order.

To obtain them, we defined these two order parameters as follows.  According to the group theory 
analysis~\cite{cb}, ${\bm H}({\bm k}=0)$ has two 1D irreducible representations which have the 
same index table and which correspond to 
the lowest eigenstate of ${\bm H}({\bm k}=0)$ in the Z$_2$ SB phase region. Let us call 
their bases as $|\phi^{(1)}_{3}\rangle$ and $|\phi^{(2)}_{3}\rangle$ respectively. 
Note also that ${\bm H}({\bm k}=0)$ has three 2D irreducible representations 
which have the same index table and which correspond to 
the doubly degenerate lowest eigenstates 
of ${\bm H}({\bm k}=0)$ in the Z$_6$ SB phase region. Call respective 
three sets of doubly degenerate bases as 
$|\phi^{(l)}_{1}\rangle$ and $|\phi^{(l)}_{2}\rangle$ ($l=1,2,3$). 

The order parameter for the Z$_2$ SB phase is given by a linear combination of 
the following two quantities;  
\begin{eqnarray}
t_{m} = \sum_{j,\alpha} S_{j,\alpha} \langle j,\alpha | \phi^{(m)}_{3} \rangle \nn 
\end{eqnarray}
with $m=1,2$ and $j$ sublattice index, $\alpha$ spin index. 
In Fig.~\ref{fig13}, the temperature dependence of either $t_1$ or $t_2$ is shown 
as the Z$_2$ SB order parameter. The XY order parameter for the Z$_6$ SB phase 
is given by a combination of the following three quantities; 
\begin{eqnarray}
m^{(l)} \equiv \sqrt{(m^{(l)}_1)^2 + (m^{(l)}_2)^2} \nn 
\end{eqnarray}
with $l=1,2,3$ and 
\begin{eqnarray}
m^{(l)}_{m} \equiv \sum_{j,\alpha} S_{j,\alpha} \langle j,\alpha | \phi^{(l)}_{m} \rangle \nn 
\end{eqnarray}
with $m=1,2$. In Fig.~\ref{fig13}, the temperature dependence of one of $m^{(1)}$, $m^{(2)}$ and 
$m^{(3)}$ is shown as the XY order parameter for the Z$_6$ SB phase.

\end{document}